\documentstyle[12pt]{article}
\makeatletter
\gdef\@pubnumber{\null}
\long\def\pubnumber#1{\gdef\@pubnumber{DAMTP 93-50}}
\def\@makepub{\vbox to \z@{\hbox to
\textwidth{\hfill\llap{\parbox[t]{0.33\textwidth}{\raggedleft\@pubnumber}}}%
\vss}}
\def\@maketitle{\newpage
\@makepub \null
\vskip 2em \begin{center}
{\LARGE \@title \par} \vskip 1.5em {\large \lineskip 0.5em
\begin{tabular}[t]{c}\@author
\end{tabular}\par}
\vskip 1em {\large \@date} \end{center}
\par
\vskip 1.5em}

\pubnumber{1}

\def\ha{\frac{1}{2}}

\begin{document}

\title{Global Strings and the Aharonov-Bohm Effect}
\author{A.C.Davis\thanks{and King's College, Cambridge} and
	A.P.Martin,\\
	{\normalsize DAMTP,}\\
	{\normalsize{Cambridge University,}}\\
	{\normalsize{Silver St.,}}\\
	{\normalsize{Cambridge.}}\\
	{\normalsize{CB3 9EW}}\\
	{\normalsize{U.K.}}\\
	{\normalsize{England.}}}

\maketitle\begin{abstract}

When a fermion interacts with a global vortex or cosmic string a
solenoidal ``gauge'' field is induced. This results in a non-trivial
scattering cross-section. For scalars and non-relativistic fermions,
the cross-section is similar to that of Aharonov and Bohm, but with
corrections. A cosmological example is compared to one in liquid
He$^{3}$-A, and important differences are discovered.

\end{abstract}
\pagebreak

\section{Introduction}

That a particle is scattered by a solenoid, or
vortex, with a cross-section per unit length that is independent of the
solenoid radius, but depends only on the flux and particle momentum,
has long been a standard result \cite{ahb}. This, the
Aharonov-Bohm [A-B] effect, has been discussed in a number of
contexts; two of the more interesting ones being
vortices in He$^3$ \cite{kh} and cosmic strings
\cite{alfw}\cite{davis},
the dominant mechanism for energy loss from a string
network, in the friction dominated era of the
early universe, arising from its Aharonov-Bohm interaction with the
surrounding plasma \cite{rohm}.

The
A-B effect was, until recently,
associated with the vector potential on the solenoid, or
a string arising when a local symmetry is broken. However, for
non-relativistic particles,
March-Russell, Preskill and Wilczek \cite{mrpw} have produced a {\em global}
analogue to the A-B effect.
They have shown that the breakdown of a
global symmetry to a discrete subgroup can lead to
particles scattering off a global vortex with an Aharonov-Bohm-like
cross-section, provided the momentum-transfer is below a certain limit.
The result obtained for the scattering amplitude, in this case, is
	\begin{equation}
		f(\phi)=
		\frac{e^{\frac{-i\phi}{2}}}{(2\pi ik)^{\frac{1}{2}}}
		\left(
		\frac{1}{\cos{(\frac{\phi}{2})}}+
		2\sum_{n=0}^{\infty} (-1)^{n}
		(e^{i\Delta_n}-1)
		\cos{((n+\frac{1}{2})\phi)}
		\right)
		\label{bong}
	\end{equation}
where
	\begin{equation}
		\Delta_n =
		\pi\left[ n+\frac{1}{2}
		-\left( \left(n+\frac{1}{2}\right)^2 +
		\frac{1}{4}\right) ^{\frac{1}{2}} \right]
	\end{equation}
$k$ is the momentum of the ingoing state and $\phi$ is the usual
azimuthal angle.
The first term in (\ref{bong}) is the usual maximal Aharonov-Bohm
amplitude. This form of $f$ gives a differential cross-section,
expressed in terms
of the scattering angle $\theta=\pi-\phi$, of
	\begin{equation}
		\frac{d\sigma}{d\theta}=
		\frac{1}{2\pi k \sin^2{\frac{\theta}{2}}}
		[1+C(\theta)]
		\label{cow}
	\end{equation}
which is the maximal Aharonov-Bohm cross-section multiplied by a calculable
correction factor, $[1+C(\theta)]$, which approaches 1 at
small angles\footnote{It has been
calculated numerically that the correction is largest (C=0.202) at
$\theta=\pi$.}.

All these calculations were done, however, using quantum mechanical
methods and \\
Schr\"{o}dingers equation, and only work in the case of suitably small
momentum transfer. Since global strings arise in many field theory
examples it is pertinent to ask how far the result extends. In
particular, for global strings relevant to particle physics and
cosmology, it is clearly necessary to generalise to the relativistic
case, and to see if the effect is confined to this particular model or occurs
more generally.
Another interesting question concerns the link between string defects in
ordered media and cosmological models. It has already been shown that
substances such as liquid crystal and superfluid helium can be used to
verify the Kibble mechanism for the formation of global strings in the
laboratory\cite{bow}\cite{chu}\cite{chu2}.
As yet, though, there are no examples of ordered media
with local symmetries, and so it is currently
impossible to physically simulate the evolution of gauge strings.
A paper by Khazan \cite{kh}, however, has postulated
the existence of an Aharonov-Bohm effect associated with
certain line defects in one of the ``modes'' of liquid He$^{3}$-A
\cite{kh}.
If there is as
strong a link between cosmological defects and those in ordered media
as we would like to think, then there is a  question whether we have
actually found an ordered media with a gauge symmetry, or whether the
Aharonov-Bohm effect is not, as March-Russell et al suggested, confined
to the case of local strings. This paper looks at the strength of the
link between March-Russell et al's cosmological model, and that in He$^{3}$-A.

In section 2 we review the results of \cite{mrpw} and show how
the A-B cross-section arises there. We also consider the
vortices found in He$^{3}$ \cite{kh}, and find that, for half-quanta
flux, there is a
correction term to the Aharonov-Bohm cross-section, not previously
discussed. In section 3 we
consider relativistic scalar particles, and show that they too exhibit an A-B
cross-section, once more subject to corrections and limitations.
In section 4 we discuss the case of integer ``induced''
string flux, and produce the corresponding analogue of \cite{mrpw},
but with corrections. We also discuss integer flux vortices in
He$^3$-A, and find that here the correction term is absent.
Our conclusions, and the possible relevance of
our work, are discussed in section 5.

\section{Non-relativistic Incidences of the Aharonov-Bohm Effect}

Consider the case of a model with a global U(1) symmetry, broken down
to $Z_{2}$ by the condensation of a scalar field
$\Phi\rightarrow\eta e^{i\phi}$. Let this
scalar field interact with a complex field $\psi$ via the coupling
\begin{displaymath}
	\Delta{\cal{L}}=g\Phi\psi^{2}+H.c.
\end{displaymath}
This is the ``frame-dragging'' model discussed in \cite{mrpw}. We now perform a
transformation
\begin{equation}
	\left(
	\begin{array}{c}
	\rho_{1}\\
	\rho_{2}\\
	\end{array}
	\right)
	=\frac{1}{\sqrt{2}}
	\left(
	\begin{array}{cc}
	e^{i\phi/2} & e^{-i\phi/2}\\
	-ie^{i\phi/2} & ie^{-i\phi/2}\\
	\end{array}
	\right)
	\left(
	\begin{array}{c}
	\psi\\
	\psi^{*}\\
	\end{array}
	\right)
\end{equation}
to obtain the Lagrangian in terms of the mass eigenstates $\rho_{1}$,
$\rho_{2}$. The corresponding masses $\mu_{1}$,$\mu_{2}$ are given by
$\mu_{2}^{1}=\sqrt{m^{2}\pm\Gamma}$, where $\Gamma=2g\eta$, so the
non-zero expectation value of $\Phi$ is seen to generate a mass
splitting between the two mass eigenstates.
Another effect is that it produces a non-zero
mixing term between the two states. These appear as the off-diagonal
terms in the equation of motion
\begin{equation}
	i\partial_{t}
	\left(
	\begin{array}{c}
	\rho_{1}\\
	\rho_{2}\\
	\end{array}
	\right)=
	\left(
	\begin{array}{cc}
	-\frac{1}{2\mu_{1}}
	\left(
	\bigtriangledown^{2}-\frac{1}{4r^{2}}
	\right)
	+\mu_{1} &
	-\frac{\partial_{\phi}}{2\mu_{1}r^{2}}\\
	\frac{\partial_{\phi}}{2\mu_{2}r^{2}} &
	-\frac{1}{2\mu_{2}}
	\left(
	\bigtriangledown^{2}-\frac{1}{4r^{2}}
	\right)
	+\mu_{2}\\
	\end{array}
	\right)
	\left(
	\begin{array}{c}
	\rho_{1}\\
	\rho_{2}\\
	\end{array}
	\right)
	\label{moog}
\end{equation}
March-Russell et al \cite{mrpw} claim that the effect of
this is ignorable provided we impose the restriction
$4k^{2}\sin^{2}{(\theta/2)}\ll\Gamma$ on the momentum transfer.
This is backed up with the suggestion
that since the two states are of differing masses, it is unlikely that
they will interact for low-incident momenta.

The transformation has
other, more important, effects however.
These are to  impose a boundary condition on $\rho$ such that
$\rho(\phi+2\pi)=-\rho(\phi)$, and to generate the additional
potential $1/4r^{2}$.
The first of these means that a partial wave solution should comprise
of half odd integer modes only. In particular, it means that, if we
ignore the effects of the additional potential for the time being, the
solution to our equations of motion involves Bessel functions of
order $\nu=n+\ha$, where $n\in Z$. In the case of a pure $\rho_{2}$
ingoing state, this would lead to the full
Aharonov-Bohm cross-section
\begin{displaymath}
	\frac{d\sigma}{d\theta}=
	\frac{1}{2\pi k}
	\frac{1}{\sin^{2}{(\theta/2)}}
\end{displaymath}
However, the additional potential
modifies this, such that the Bessels functions are, instead, of order
$\nu=\sqrt{(n+\ha)^{2}+\frac{1}{4}}$. The effect of this is to make
the actual cross-section that given in (\ref{cow}).

In \cite{mrpw}, Khazan's paper on superfluid
He$^{3}$-A \cite{kh} is cited as another possible example of a model
with broken global symmetry exhibiting an Aharonov-Bohm cross-section,
though the authors of \cite{mrpw} say that they are unsure
whether this effect ``falls
into [their] framework''. The reasons for this are easily seen when
one examines Khazans model, since it displays both important
similarities {\em and} differences to \cite{mrpw}. One of the points
that worried them, however, the absence of an additional potential, is
erroneous.

Liquid helium differs from the fields involved in
\cite{mrpw} in that it has a {\em matrix} order parameter, $A_{\alpha i}$.
In the natural state this has symmetry group \cite{volkh}
\begin{displaymath}
	G=SO(3)^{spin}\times SO(3)^{orb}\times U(1) \times Z_{2}
\end{displaymath}
where SO(3)$^{spin}$ corresponds to three-dimensional rotations under
which the first (spin) index of $A_{\alpha i}$ transforms as a
vector, SO(3)$^{orb}$ corresponds to similar rotations of the second
(orbital) index, U(1) consists of transformations taking $A_{\alpha
i}\rightarrow A_{\alpha i}e^{i\beta}$ and Z$_{2}$ consists of the two
elements 1 and T, where T is the operation of time reversal. In the
superfluid A-phase the symmetry is reduced and it is possible
to write the order parameter in the form
\begin{displaymath}
	A^{0}_{\alpha i}=
	d_{\alpha}(\underline{r})
	(\Delta^{\prime}_{i}(\underline{r})+
	i\Delta^{\prime\prime}_{i}(\underline{r}))
\end{displaymath}
where $d_{\alpha}$, $\Delta^{\prime}_{i}$ and
$\Delta^{\prime\prime}_{i}$ are mutually orthogonal unit vectors.
It retains, however, two combined symmetries \cite{volsal}.

The first of these is a discrete symmetry under
$\underline{d}\rightarrow-\underline{d}$,
A$_{\alpha i}\rightarrow$A$_{\alpha i}e^{i\pi}$.
This is reflected in the factorisation by Z$_{2}$ of the corresponding
order parameter space
\begin{displaymath}
	R=(S^{2}\times SO(3))/Z_{2}
\end{displaymath}

The second combined symmetry is a continuous one.
Defining the angular momentum of the system by
$\underline{l}=
\underline{\Delta}^{\prime}\times\underline{\Delta}^{\prime\prime}$, this
corresponds to rotations about $\underline{l}$ by an arbitrary angle
$\beta$ coupled to a multiplication of the order parameter by
$e^{i\beta}$.

We now consider the case where this order
parameter oscillates with respect to its equilibrium value $A^{0}_{\alpha i}$
such that $A_{\alpha i}=A^{0}_{\alpha i}+\delta A_{\alpha i}$, where
$\delta A_{\alpha i}=\psi
d_{\alpha}(\Delta^{\prime}_{i}-i\Delta^{\prime\prime}_{i})$ and $\psi$
is a complex scalar describing what is termed the ``clapping'' mode
\cite{volkh}.
Consideration of the first homotopy group of the space of degenerate
states, R,  shows that
\begin{displaymath}
	\pi_{1}(R)=Z_{4}
\end{displaymath}
from which we see that this model supports four different topological
classes of linear defects, or vortices \cite{volmin}. We consider the
class with
quanta $\ha$. Near such a vortex, the order parameter can be written
as
\begin{displaymath}
	\begin{array}{rcl}
	d_{\alpha} & = &
	\hat{x}_{\alpha} \cos{(\phi/2)}-
	\hat{y}_{\alpha} \sin{(\phi/2)}\\
	 & & \\
	\Delta^{\prime}_{i}+
	i\Delta^{\prime\prime}_{i} & = &
	(\hat{x}_{i}+i\hat{y}_{i})
	e^{i\phi/2}\\
	\end{array}
\end{displaymath}
where $\phi$ is the azimuthal angle in the plane, and $\underline{\hat{x}}$,
$\underline{\hat{y}}$ are the usual unit vectors. It is the interaction of the
``clapping mode'' with this vortex that gives rise to the
Aharonov-Bohm cross-section.

The equation of motion for such a system
is found to be
\begin{displaymath}
	c^{2}({\bf \bigtriangledown}-i{\bf A})^{2}\psi +U\psi=
	(\omega_{0}^{2}-\omega^{2})\psi
\end{displaymath}
where
${\bf A}=\Delta^{\prime}_{i}{\bf \bigtriangledown}\Delta^{\prime\prime}_{i}$
and
$U=-c^{2}(\bigtriangledown_{k}d_{\alpha})^{2}$.

Using the form of the vortex given above one can write these more
explicitly as
\begin{equation}
	\begin{array}{ccc}
	U=\frac{1}{4r^{2}} & , &
	{\bf A}=\frac{1}{2r}\hat{\bf {e}}_{\phi}\\
	\end{array}
\end{equation}
Not only is the ``gauge'' field very similar to the effective
``gauge'' field  one obtains as a result of the transformation in
\cite{mrpw}, but there is also an identical additional potential. At
first glance then, it would appear that the superfluid helium model
and the ``frame-dragging'' case are very closely linked.
There are some important differences however.

In both \cite{mrpw} and \cite{kh} the equation of motion involves the
square of a derivative term, the only difference
between the two cases in this term,arising in the contribution of the azimuthal
component.
In \cite{mrpw} the relevant term is
\begin{displaymath}
	\left[\frac{1}{r^{2}}\partial_{\phi}^{2}-
	\frac{2i\sigma_{2}}{r}\left(\frac{1}{2r}\right)\partial_{\phi}-
	\left(\frac{1}{2r}\right)^{2}\right]\psi
\end{displaymath}
where the $\frac{1}{2r}$ is the induced ``gauge field'', and
$\sigma_{2}$ is the second Pauli matrix.
As explained above, a partial wave solution must comprise of
only half odd integer modes, $e^{i(n+\ha)\phi}$, so we see that the first
term in brackets is equivalent to $-(n+\ha)^{2}\psi$.
This together with the third
term determine the order of the Bessel functions
$\nu^{2}=(n+\ha)^{2}+\frac{1}{4}$. The second term is none other than
the off-diagonal terms, dismissed in \cite{mrpw} for suitably low
momentum transfer.

In \cite{kh} the corresponding term is
\begin{displaymath}
	\left[\frac{1}{r^{2}}\partial_{\phi}^{2}-
	\frac{2i}{r}\left(\frac{1}{2r}\right)\partial_{\phi}-
	\left(\frac{1}{2r}\right)^{2}\right]\psi
\end{displaymath}
where once more the $\frac{1}{2r}$ is due to the ``gauge'' potential.
Here, however, we have no strange boundary condition, so using a
partial wave solution with integer modes, $e^{in\phi}$,  we see that the whole
term is
equivalent to $-(n+\ha)^{2}\psi$. We also note that there are
no off-diagonal terms.
Hence, the result in \cite{kh} is unconstrained by the
momentum transfer limit in \cite{mrpw}. We still have an
extra potential, $U=1/4r^{2}$, however, and
this means that the Bessel functions in the solution will be of order
$\nu^{2}=(n+\ha)^{2}+\frac{1}{4}$, exactly the same as in \cite{mrpw}.
Hence, Khazans stated cross-section is not quite the full result, as it
does not include corrections due to the potential $U$. If we include
this effect, then the cross-section is identical to that in
\cite{mrpw}.

At first sight, this would appear to add more weight to the idea of
the two being different incarnations of the same phenomenon. However,
it is important to remember that although they share the same result,
they possess it for quite different reasons.
The additional potential in \cite{mrpw}
arises as a direct result of the induced ``gauge field'', whilst the
Aharonov-Bohm effect is produced by the imposed boundary condition. In
the case of superfluid He$^{3}$-A, however, things are a little more
complicated; there are in fact {\em two} defects present. The string
solution is actually contained in the vortex form of
$\Delta^{\prime}+i\Delta^{\prime\prime}$, whilst the similar form of ${\bf
d}$ is actually the superposition of a disclination of the ${\bf d}$
field upon the string solution \cite{volmin}\cite{salvo}.
It is this latter defect which gives rise to
the additional potential. Unlike the ``frame-dragging'' model, it is
the ``gauge'' field which causes the Aharonov-Bohm effect in superfluid
He$^{3}$-A;
just as it does
in the case of local strings.

\section{Relativistic Charged Scalars}

We now consider the scattering of relativistic
charged scalars off a global string
using the following Lagrangian
\begin{equation}
{\cal{L}}=
\ha(\partial_{\mu}\rho)^{*}(\partial^{\mu}\rho)
	-\ha m^{2}\rho^{*}\rho
	-\ha g\Phi\rho^{2}
	-\ha g^{*}\Phi^{*}\rho^{*2}
	\label{lag}
\end{equation}
where $\rho$ is the scalar field and $\Phi$ is the Higgs field.
If the Higgs field condenses such that at large distances
$\Phi \rightarrow \eta e^{i\phi}$
then, on performing the transformation
$\rho \rightarrow e^{\frac{-i\phi}{2}}\rho$
and setting $A_{\mu}=\frac{1}{i}\partial_{\mu}(\frac{-i\phi}{2})$
 we get a modified Lagrangian of the form
\begin{equation}
	{\cal{L}}=
	\ha(D_{\mu}\rho)^{*}
	(D^{\mu}\rho)
	-\ha m^{2}\rho^{*}\rho
	-\ha v(g\rho^{2}+g^{*}\rho^{*2})
\end{equation}
with induced ``gauge'' field
$A_{\phi}=-\ha$, $A_{t,r,z}=0$,
where
$D_{\mu}=\partial_{\mu}+iA_{\mu}$.
A simple application of the Euler-Lagrange equations now yields the
equation of motions for $\rho$,$\rho^{*}$.
\begin{equation}
	\begin{array}{lllllll}
	D_{\mu}D^{\mu}\rho & + & m^{2}\rho
	& + & 2vg^{*}\rho^{*} & = & 0 \\
	D_{\mu}^{*}D^{\mu*}\rho^{*} & + & m^{2}\rho^{*}
	& + & 2vg\rho & = & 0 \\
	\end{array}
	\label{mot}
\end{equation}
It is possible to write (\ref{mot}) in matrix form by setting
\begin{equation}
	{\cal{D}}_{\mu}{\cal{D}}^{\mu}=
		\left(
		\begin{array}{cc}
		D_{\mu}D^{\mu} & 0 \\
		0 & D_{\mu}^{*}D^{\mu*} \\
		\end{array}
		\right)
	,
	\tilde{\rho}=
		\left(
		\begin{array}{c}
		\rho \\
		\rho^{*} \\
		\end{array}
		\right)
	,
	{\cal{M}}^{2}=
		\left(
		\begin{array}{cc}
		m^{2} & 2vg^{*} \\
		2vg & m^{2} \\
		\end{array}
		\right)
\end{equation}
whereupon (\ref{mot}) becomes
\begin{equation}
	{\cal{D}}_{\mu}{\cal{D}}^{\mu}\tilde{\rho}+{\cal{M}}^{2}\tilde{\rho}=0
	\label{matty}
\end{equation}
We next attempt to decouple the equations for  $\rho$ and $\rho^{*}$ by
diagonalizing ${\cal{M}}^{2}$.To do this we first need to find its
eigen-values and eigen-states. These are found to
be $m^{2}\pm 2\eta|g|$
with corresponding eigen-states $\frac{1}{\sqrt{2}}(\pm\frac{g^{*}}{|g|},1)$.
If we now define
\begin{equation}
	S=\frac{1}{\sqrt{2}}
	\left(
	\begin{array}{cc}
	\alpha & -\alpha \\
	1 & 1 \\
	\end{array}
	\right)
\end{equation}
and perform the transformation
$\tilde{\rho}\rightarrow \hat{\rho}=S\tilde{\rho}$,
where
$\alpha=\frac{g^{*}}{|g|}$ (note:$|\alpha|=1$)
and
$\hat{\rho}=(\rho_{1},\rho_{2})$,
we see that
${\cal{D}}_{\mu}{\cal{D}}^{\mu}$ becomes
\begin{equation}
	\begin{array}{lcl}
	S^{\dagger}{\cal{D}}_{\mu}{\cal{D}}^{\mu}S & =	&
		\ha
		\left(
		\begin{array}{rr}
		\Delta^{2}+\Delta^{*2} & -\Delta^{2}+\Delta^{*2} \\
		-\Delta^{2}+\Delta^{*2} & \Delta^{2}+\Delta^{*2} \\
		\end{array}
		\right)
	\end{array}
\end{equation}
where $\Delta^{2}=D_{\mu}D^{\mu}$.
Using
\begin{equation}
	\begin{array}{lcl}
	\Delta^{2} & = &
	\partial_{\mu}\partial^{\mu} - A_{\mu}A^{\mu} +
	2iA_{\mu}\partial^{\mu}  \\
	\end{array}
\end{equation}
we can write (\ref{matty}) as
\begin{equation}
	\left(
	\begin{array}{cc}
	\partial_{\mu}\partial^{\mu} - A_{\mu}A^{\mu} + \mu_{1}^{2} &
	-2iA_{\mu}\partial^{\mu}  \\
	-2iA_{\mu}\partial^{\mu}  &
	\partial_{\mu}\partial^{\mu} - A_{\mu}A^{\mu} + \mu_{2}^{2} \\
	\end{array}
	\right)
	\left(
	\begin{array}{c}
	\rho_{1} \\
	\rho_{2} \\
	\end{array}
	\right)
	=0
\end{equation}
where $(\mu_{2}^{1})^{2}=m^{2}\pm2|g|v$.
This is a similar form to that in \cite{mrpw}, though here we have a
Klein-Gordon rather than a Schr\"{o}dinger equation.

Proceeding in a similar fashion to \cite{mrpw}
we now ignore off-diagonal
terms, and, using cylindrical polars and assuming that
all motion takes place in
the ($r$,$\phi$) plane we obtain
\begin{equation}
	\left(
	\begin{array}{cc}
	\partial_{t}^{2} - {\cal{5}}^{2} + \frac{1}{4r^{2}} + \mu_{1}^{2}
	& 0 \\
	\partial_{t}^{2} - {\cal{5}}^{2} + \frac{1}{4r^{2}} + \mu_{2}^{2}
	& 0 \\
	\end{array}
	\right)
	\left(
	\begin{array}{c}
	\rho_{1} \\
	\rho_{2} \\
	\end{array}
	\right)
	=0
\end{equation}
We now take the case of a pure $\rho_{2}$ ingoing
state.
\begin{equation}
	\begin{array}{rcl}
	-\frac{1}{2\mu_{2}}\partial_{t}^{2}\rho_{2} & = &
	[-\frac{1}{2\mu_{2}}({\cal{5}}^{2} - \frac{1}{4r^{2}})
	+ \frac{\mu_{2}}{2}]\rho_{2} \\
	\end{array}
	\label{loop}
\end{equation}
So, by modification of the solution in \cite{mrpw}, we find that
\begin{equation}
	\rho_{2}(t,r,\phi)=
	\sum_{n\in Z}
	e^{-i(\omega +\frac{\mu_{2}}{2})^{\ha}\sqrt{2\mu_{2}}t}
	e^{i(n+\ha)\phi}P_{n}^{(2)}(r)
\end{equation}
Hence, following \cite{mrpw} we obtain the
following differential cross-section
\begin{equation}
	\frac{d\sigma}{d\theta}=
	\frac{1}{2\pi k_{2}\sin^{2}(\frac{\theta}{2})}[1+C(\theta)]
\end{equation}
where $C(\theta)$ is the same correction as given earlier, and in \cite{mrpw}.
However, we have used the same assumption as \cite{mrpw} in ignoring
the off-diagonal terms, so this result, though relativistic, is
constrained by the same requirement on the momentum transfer as \cite{mrpw}.

\section{Effective Integer Flux in Non-Relativistic Case}

Now consider the case where $\Phi$ condenses to
$\eta e^{2i\phi}$ instead of $\eta e^{-i\phi}$. The result of this
is to remove the extra boundary condition and modify the
additional potential to $1/r^{2}$. This time then, the allowed spectrum
of partial waves includes only integers, and it is easy to obtain a
solution by modifying that found in \cite{mrpw}. Considering once more
a pure $\rho_{2}$ ingoing state we have
\begin{equation}
	\rho_{2}(t,r,\phi)=
	\sum_{n\in Z}
	e^{-i(\omega+\mu_{2})t}e^{in\phi}P_{n}^{(2)}(k_{2}r)
\end{equation}
where the $P_{n}$ satisfy Bessel functions of order $\nu$ where
$\nu^{2}=(n+1)^{2}+1$. Since $\nu \neq 0$ for any value of $n$ we can
take $J_{\nu}$ and $N_{\nu}$ to be the two independent solutions of
the Bessel equation so that
\begin{equation}
	\rho_{2}(t,r,\phi)=
	\sum_{n\in Z}
	e^{-i(\omega+\mu_{2})t}e^{in\phi}
	\left[ a_{n}J_{\nu}(k_{2}r)+b_{n}N_{\nu}(k_{2}r)\right]
\end{equation}
We now match this onto to an incoming plane wave plus an outgoing
scattered wave at infinity such that
\begin{equation}
	\rho_{2}(t,r,\phi)\sim
	\sum_{n\in Z}
	e^{-i(\omega+\mu_{2})t}\left[ e^{in\phi}e^{-i\pi |n|/2}
	J_{|n|}(k_{2}r)+
	\frac{e^{ik_{2}r}}{\sqrt{r}}f_{n}e^{in\phi}\right]
\end{equation}
where we have made use of the usual expansion of the plane wave in
terms of integer order Bessel functions
\begin{equation}
	e^{-ikr\cos{\phi}}=
	\sum_{n\in Z}e^{-i\pi |n|/2}e^{i|n|\phi}J_{|n|}(kr)
\end{equation}
If we now set $z=k_{2}r$ and make use of the asymptotic forms of the
Bessel functions
\begin{eqnarray}
	J_{\mu}(x) &\simeq & \sqrt{\frac{2}{\pi x}}
	\cos{(x-\frac{\mu\pi}{2}-\frac{\pi}{4})}\\
	N_{\mu}(x) &\simeq & \sqrt{\frac{2}{\pi x}}
	\sin{(x-\frac{\mu\pi}{2}-\frac{\pi}{4})}\\
	\nonumber
	\label{ping}
\end{eqnarray}
then taking incoming and outgoing components (corresponding to
$e^{-ikr}$ and $e^{ikr}$ respectively) we obtain
\begin{equation}
	\begin{array}{rcrcrcr}
	\frac{1}{\sqrt{2\pi k}}e^{i\nu\pi/2+i\pi/4}a_{n} & + &
	i\frac{1}{\sqrt{2\pi k}}e^{i\nu\pi/2+i\pi/4}b_{n} & = &
	\frac{1}{\sqrt{2\pi k}}e^{i\pi/4} & & \\
	\frac{1}{\sqrt{2\pi k}}e^{-i\nu\pi/2-i\pi/4}a_{n} & - &
	i\frac{1}{\sqrt{2\pi k}}e^{-i\nu\pi/2-i\pi/4}b_{n} & = &
	\frac{1}{\sqrt{2\pi k}}e^{-i\pi|n|-i\pi/4} & + & f_{n}\\
	\end{array}
\end{equation}
from which we can extract an expression for $f_{n}$ in terms of
$b_{n}$
\begin{equation}
	f_{n}=\frac{1}{\sqrt{2\pi ik}}
	\left[
	e^{-i\nu\pi}-e^{-i\pi|n|}-2ie^{-i\nu\pi/2}b_{n}
	\right]
\end{equation}
If, for the time being, we ignore the $b_{n}$, then we have
\begin{equation}
	f(\phi)=\sum_{n\in Z} f_{n}e^{in\phi}=
	\sum_{n\in Z}
	\frac{1}{\sqrt{2\pi ik}}
	\left[
	e^{-i\nu\pi}-e^{-i\pi|n|}
	\right]
	e^{in\phi}
\end{equation}
By rearrangement of the series, it is possible to rewrite this as
\begin{equation}
	f(\phi)=
	\frac{1}{\sqrt{2\pi ik}}
	\sum_{n=0}^{\infty}
	2(-1)^{n+1}(e^{i\Delta_{n}}+1)e^{-i\phi}\cos{((n+1)\phi)}
\end{equation}
c.f. the correction term in \cite{mrpw}. This comes as no surprise
since the effective ``induced'' flux in this case is $1$ and since
$\sin{(\pi)}=0$ we expect the Aharonov-Bohm part of the scattering
amplitude to vanish leaving us with the correction alone.
Hence, to leading order, the scattering cross-section is
entirely a result of the induced $1/r^{2}$ potential.
The next step is to determine what effect the $b_{n}$ may have on
the scattering.

In considering the r\^{o}le played by the $b_{n}$ it is necessary to
consider the solution inside the core, as we can then get an
expression for $b_{n}$ by matching the internal and external solutions
at the core radius. Demanding regularity and square-integrability at
the origin we obtain
\begin{equation}
	\sum_{n\in Z} c_{n}J_{n}(k^{\prime}_{2}r)e^{in\phi}
\end{equation}
as our internal solution, whilst, from above, our external solution is
\begin{equation}
	\sum_{n\in Z}
	\left[ a_{n}J_{\nu}(k_{2}r)+b_{n}N_{\nu}(k_{2}r)\right]e^{in\phi}
\end{equation}
It is important to note that $k^{\prime}_{2}$ and $k_{2}$ are different due to
the change in mass on passing into the core.
Assuming that $k_{2}^{(\prime)}r\ll 1$ we can match at $r=R$,
making use of the small
argument form of the Bessel functions
\begin{equation}
	\begin{array}{ccc}
	J_{\mu}(x)\simeq\left(\frac{x}{2}\right)^{\mu}\frac{1}{\mu!} &
	, &
	N_{\mu}(x)\simeq\-\frac{(\mu-1)!}{\pi}
	\left(
	\frac{2}{x}
	\right)^{\mu}\\
	\end{array}
	\label{pang}
\end{equation}
The matching conditions involved for Schr\"{o}dinger's equation are
discussed in \cite{amrw} where the solution is shown to be continuous
up to and including it's first derivative. Using this, and defining
$z=k_{2}r$, $z^{\prime}=k_{2}^{\prime}r$ we obtain
\begin{equation}
	\begin{array}{rcrcr}
	J_{\nu}(z^{\prime}) & = &
	J_{\nu}(z)a_{n} & + &
	N_{\nu}(z)b_{n} \\
	J_{\nu}^{\prime}(z^{\prime}) & = &
	J_{\nu}^{\prime}(z)a_{n} & + &
	N_{\nu}^{\prime}(z)b_{n} \\
	\end{array}
\end{equation}
where $^{\prime}$ denotes $\frac{d}{dr}$. Combining these gives
\begin{equation}
	\frac{b_{n}}{a_{n}}=
	\frac{J_{\nu}(z)J_{\nu}^{\prime}(\tilde{z})-
	J_{\nu}(\tilde{z})J_{\nu}^{\prime}(z)}
	{N_{\nu}(z)J_{\nu}^{\prime}(\tilde{z})-
	J_{\nu}(\tilde{z})N_{\nu}^{\prime}(z)}
	\label{woof}
\end{equation}
and we can now make use of the Bessel function identity
\begin{equation}
	\begin{array}{ccc}
	\frac{d}{dr}J_{\mu}(r) & = &
	\frac{1}{2}(J_{\mu-1}(r)-J_{\mu+1}(r))\\
	\end{array}
	\label{bess}
\end{equation}
and their small argument forms to reduce (\ref{woof}) to
\begin{equation}
	\frac{b_{n}}{a_{n}}\simeq
	(kR)^{2\nu}
\end{equation}
Since $\nu=[(n+1)^{2}+1]^{\ha}$, we see that the relative suppression
of $b_{n}$ to $a_{n}$ is never less than $kR$. If however we were to
remove the induced potential such that $\nu=n+1$ then for $n=-1$,
$\nu$ would be zero, and
\begin{equation}
	\frac{b_{-1}}{a_{-1}}\simeq
	\frac{\pi}{2\log{(kR)}}
\end{equation}
so that for this mode $b_{-1}$ would actually {\em dominate} $a_{-1}$,
and we would recover Everetts cross-section for scattering off a
local-string of integer flux \cite{eve}:
\begin{displaymath}
	\frac{d\sigma}{d\phi}=
	\frac{\pi}{4k}\frac{1}{[\log{(kR)}]^{2}}.
\end{displaymath}
This is what, in fact, happens
in the case of liquid helium.

If we consider the He$^{3}$-A string in the case of
integer flux then we find that the vortex solution slightly changes,
in that there is now no disclination present in the spin field,
and $d_{\alpha}$ is found to be constant. Since
$U=-c^{2}(\bigtriangledown_{k}d_{\alpha})^{2}$, this implies that the
additional potential is zero. Hence, unlike the ``frame-dragging''
model in this instance, we obtain the {\em uncorrected} Everett
cross-section, since the leading order correction term is also absent.

\section{Conclusions}

We have seen then that it is possible to construct a model where a
global string exhibits an Aharonov-Bohm-like cross-section on both a
non-relativistic and relativistic level - for scalars. It
should be noted, however,
that the relativistic model is subject to even stronger
restrictions than the non-relativistic one, and so we cannot say
much about the general relativistic case. This may
not, though, necessarily be the case for relativistic fermions.

The case of He$^{3}$-A is, however, somewhat different. Firstly,
the defect involved is, essentially, an {\em half-integer} ``flux'' string -
as opposed to the integer ``flux'' string of March-Russell et al.
Secondly, there is a difference in the source of
the additional potential; that of \cite{mrpw} coming straight from the
``gauge field'', and that in \cite{kh} being the result of a {\em
second} defect, superimposed on the first. Hence, in the case of
integer flux, where He$^{3}$-A supports no such defect, the additional
potential vanishes - in contrast to the everpresent potential of
\cite{mrpw}. Finally, the Aharonov-Bohm cross-section itself is
attributable to different sources in the two models. In the
``frame-dragging'' case it is essentially a result of the second
order nature of the equation of motion following a transformation,
whilst superfluid He$^{3}$-A demonstrates Aharonov-Bohm scattering for
the same reason a local string does -
the presence of a long-range $1/2r$ potential.

In \cite{kh} it was suggested that the A-B cross-section of the
half-integer ``flux'' string in He$^{3}$-A might allow it to be distinguished
experimentally from an integer ``flux'' string. However, since the two
cross-sections only differ from each other by a log($kR$) factor, it
seems unlikely that any experiment would be sensitive enough to
distinguish between these two cross-sections. However, in section 2 we
saw that there is a correction, $C(\theta)$, to the A-B cross-section
in the case of the half-integer ``flux'' string, something missed in
\cite{kh}. The angular dependence of this factor may allow the two
vortices to be distinguished experimentally in a scattering
experiment.

One of us (A.P.M.) acknowledges the S.E.R.C. for financial support,
and would like to thank R.Rohm for communicating ref.\cite{rohm} and
correspondence.  We would
like to thank R.Brandenberger, A.Goldhaber and M.Hindmarsh for
discussions.


\begin{thebibliography}{99}
	\bibitem{ahb} Y.Aharonov and D.Bohm {\em Phys.Rev.} ${\bf
		115} (1959), 485$
	\bibitem{kh} M.V.Khazan {\em Pis'ma Zh.Eksp.Teor.Fiz.} ${\bf
		41} (1985), 396$
	\bibitem{alfw} M.G.Alford and F.Wilczek {\em Phys.Rev.Lett.}
		${\bf 62} (1989), 1071$
	\bibitem{davis} L.Perivolaropoulos, A.Matheson, A.C.Davis and
		R.H.Brandenberger \\
		{\em Phys.Lett.} ${\bf B245} (1990), 556$
	\bibitem{rohm} R.Rohm {\em Ph.D. thesis}
	\bibitem {mrpw} J.March-Russell, J.Preskill and F.Wilczek
		{\em Phys.Rev.Lett.} ${\bf 68} (1992), 2567$
	\bibitem{bow} M.J.Bowick, L.Chandar, E.A.Schiff and
		A.M.Srivastava, {\em SU-HEP-}4241{\em -}512,
		{\em TPI-MINN-}92135-T
	\bibitem{chu} I.Chuang, R.Durrer, N.Turok and B.Yurke {\em
		Science} ${\bf 251}(1991), 1336$, B.Yurke,
		A.N.Pargellis, I.Chuang and N.Turok, {\em PUPT-}91{\em -}1274
	\bibitem{chu2} I.Chuang, N.Turok and B.Yurke {\em
		Phys.Rev.Lett.} ${\bf 66}(1991), 2472$
	\bibitem{eve} A.E.Everett {\em Phys.Rev.D} ${\bf 24} (1981),
		858$
	\bibitem{amrw} M.Alford, J.March-Russell and F.Wilczek {\em
		Nuc.Phys.} ${\bf B328}(1989),140$
	\bibitem{volkh} G.E.Volovik and M.V.Khazan {\em Sov.Phys.JETP}
		${\bf 58}(1983),551$
	\bibitem{volsal} G.E.Volovik and M.M.Salomaa {\em
		Sov.Phys.JETP} ${\bf 61}(1985),986$
	\bibitem{volmin} G.E.Volovik and V.P.Mineev {\em Sov.Phys.JETP}
		${\bf 45}(1977),1186$
	\bibitem{salvo} M.M.Salomaa and G.E.Volovik {\em
		Rev.Mod.Phys.} ${\bf 59}(1987),533$
	\bibitem{mnee} A-C.Davis, A.P.Martin and N.Ganoulis {\em DAMTP 93-46}
\end{thebibliography}
\end{document}